\documentclass{ws-ijmpd}
\usepackage[super,compress]{cite}
\usepackage{hyperref}
\usepackage{color}
\usepackage{newunicodechar}

\newcommand{\Ms}{{\ensuremath{{M}_{\odot} }}}

\newcommand{\Cx}{{\ensuremath{^{12}\mathrm{C}}}}
\newcommand{\Ni}{{\ensuremath{^{56}\mathrm{Ni}}}}
\newcommand{\Fe}{{\ensuremath{^{56}\mathrm{Fe}}}}

\newcommand{\Ox}{{\ensuremath{^{16}\mathrm{O}}}}

\newcommand{\KEPLER}{\texttt{KEPLER}}
\newcommand{\STELLA}{\texttt{STELLA}}

\newcommand{\erg}{{\ensuremath{\mathrm{erg}}}}

\newcommand{\Msun}{\ensuremath{\mathrm{M}_\odot}}

\newcommand{\note}[1]{\emph{\textcolor{red}{}}}

\newcommand{\citep}[1]{{(\refcite{#1})}}
\newcommand{\citet}[1]{{(\refcite{#1})}}

\newcommand{\araa}{ARA\&A}%
\newcommand{\apj}{ApJ}%
\newcommand{\apjl}{{ApJ}}%
\newcommand{\aap}{{A\&A}}%
\newcommand{\mnras}{{MNRAS}}%
\newcommand{\pasp}{{PASP}}%
\newcommand{\ssr}{{Space~Sci.~Rev.}}%
\newcommand{\nat}{{Nature}}%
\newcommand{\pasa}{{Publ. Astron. Soc. Austral.}}%

\begin{document}

\markboth{Ke-Jung Chen}
{Superluminous Supernovae}

%
\catchline{}{}{}{}{}
%

\title{Physics of Superluminous Supernovae}

\author{Ke-Jung Chen}

\address{Institute of Astronomy and Astrophysics, Academia Sinica, Taipei 10617, Taiwan\\
kjchen@asiaa.sinica.edu.tw}


\maketitle

\begin{history}
\received{Day Month Year}
\revised{Day Month Year}
\end{history}

\begin{abstract}
Understanding how massive stars die as supernovae is a crucial question in modern astrophysics. 
Supernovae are powerful stellar explosions and key drivers in the cosmic baryonic cycles by injecting their
explosion energy and heavy elements to the interstellar medium that forms new stars.
After decades of effort, astrophysicists have built up a stand model for the 
explosion mechanism of massive stars. However, this model is challenged by new kinds of stellar explosions  discovered in the recent transit surveys. In particular,  the new population called superluminous supernovae, which are a hundred times brighter than typical supernovae, is revolutionizing our understanding of supernovae.  New studies suggest the superluminous supernovae are associated with the unusual demise of very massive stars and their extreme supernovae powered by the radioactive isotopes or compact objects formed after the explosion. Studying these supernovae fills a gap of knowledge between the death of massive stars and their explosions; furthermore, we may apply their intense luminosity to light up the distant universe. This paper aims to provide a timely review of superluminous supernovae physics, focusing on the latest development of their theoretical models.  
\end{abstract}


\ccode{PACS numbers:}


\section{Introduction}
Core-collapse (CC) supernovae (SNe), the most energetic explosions in the universe, are related to the death of massive stars \citep{bethe1990,arnett1996,woosley2002}. After a serious of nuclear fusions, a massive star of mass $ > 10 \Ms$ eventually grows an iron core at the end of its evolution. When the mass of iron core exceeds its Chandrasekhar mass \citep{chand1942} $\sim 1.5 \Ms$, electrons' degenerate pressure fails to support its own gravity, which then leads to a runaway collapse, and eventually a proto-neutron star forms. The gravitational binding energy of the iron core is released by neutrinos, which deposit a small fraction of their energy to the infalling gas and reverse the collapse into an explosion. The exploding mechanism of CC SNe contains rich explorations of fundamental science such as nuclear astrophysics, particle physics, and general relativity. Over decades of effort, both theorists and observers have gained the in-depth understanding of CC SNe by studying their explosion mechanism, nucleosynthesis, and compact remnant. In terms of stellar feedback, CC SNe are potent engines in driving the cosmic evolution by dispersing newly forged metals and injecting much energy to the interstellar medium, which seeds new star formation and powers strong galactic outflows. 

SNe emit plenty of radiation in a short time and outshines their host galaxies for weeks.  High luminosity with a short duration (a few weeks) makes SNe a powerful tool to probe the distant universe. 
Understanding SNe allows us to calculate the amount of radiation energy streaming from them; then, we can estimate the distance between the explosions and the Earth based on the brightness we observe.For example,  this method has been applied for SNe Ia to determine the expansion rate of the universe.  Because of the development of modern optical devices such as charge-coupled devices (CCDs), the detection rates of SNe increased dramatically. New transit surveys \citep{snf1,snf2,ptf1,ptf2}, have rapidly expanded the volume of the CC SNe database and sharpened our understanding of them. However, new types of astrophysical explosions observed in the past decade also challenged the stand model of CC SNe.

Discovery of superluminous SNe (SLSNe) such as SN 2006gy and SN 2007bi \citep{smith2007,quimby2007,quimby2011}, which have a peak luminosity $\sim 100$ times brighter 
than typical SNe well documented in the literature \refcite{filip1997,smartt2009}, opens a new sector in the SN classification. Their detailed observational properties can be found in excellent reviews \citep{galyam2012,Mor18,galyam19}. In general, the amount of radiation energy from SLSNe is $\sim 10^{51} \erg$, comparable to the explosion energy of CC SNe, which has a radiation energy budget of $\sim 10^{49} \erg$. The population of SLSNe is small, comprising $< 5\%$ of the total number of SNe \citep{galyam2012}, and they are usually found in dwarf galaxies. The extreme luminosity of SLSNe challenges our understanding of CC SNe. First, the luminosity of CC SNe can be approximated in the form: $\propto 4\pi R^2 T^4$, where $R$ is the radius of the photosphere, and $T$ is its effective surface temperature. If we assume the overall luminosity from the black body emission of hot ejecta, it requires either a larger $R$ or a higher $T$ to produce more luminous SNe. $R$ is determined when the hot ejecta becomes optically thin; then the photons start to stream freely. $T$ depends on the ejecta's thermal energy, which is related to the explosion or other energy inputs. The duration of light curves is associated with  the mass of the ejecta, which determines the diffusion time scale of photons and the volume of radiation reservoir. Besides the explosion energy, the decay energy of radioactive isotopes such as \Ni{} also plays a critical role in brightening SNe.  However, the explosion energy and \Ni\ production of typical CC SNe fail to explain the enormous radiation from SLSNe. Therefore, SLSNe likely originates from extreme SNe rather than typical CC SNe. 

SLSNe may hold the key to reveal the physics of exotic stellar explosions that profoundly impact  astrophysics and cosmology. Therefore, this paper aims to provide a timely review of the current understanding of SLSNe. We first discuss the physics of three promising SLSNe models and their recent development in Section 2; then summarize key issues and conclude with a future perspective  in Section 3.

\section{Theoretical Models of SLSNe}
Three kinds of emission mechanisms possibly explain the luminosity of SLSNe. The first mechanism is related to rapidly rotating neutron stars formed after the CC explosion \citep{kasen2010,woosley2010}.  A strong magnetized and rapidly spinning neutron star---magnetar can release its spin-down energy through the dipole radiation and power the luminosity of SLSNe. 
The second mechanism comes from the interaction between shocks from the stellar explosion and its surrounding circumstance-stellar medium (CSM)   \citep{chevalier1995,smith2007,moriya2010,chatz2011,chevalier2011,moriya2013}. When the SN's ejecta  runs into the medium formed before the star dies, the kinetic energy of the shock effectively converts into thermal radiation through colliding with the dense CSM, which is formed from outbursts of massive stars of $80 - 150\Msun$. The third mechanism is from the radioactive decay of \Ni. The massive stars of $150 - 250\Msun$ die as energetic thermonuclear SNe and synthesize a large amount of radioactive element, \Ni{}  up to 50 $\Ms$, and its decay energy can provide the radiation budget of SLSNe. In the following subsections, we introduce the physics of these models and their recent development.

\subsection{Magnetar-Powered SNe}
Magnetars are strongly magnetized and rapidly spinning neutron stars, with a magnetic field strength exceeding 10$^{13}$ Gauss (G) and a spin period up to a few milliseconds (ms). During the CC explosion, a rotating iron-core collapse may form into a neutron star with a strong magnetic field and rapid spin \citep{Dun92,Tho93,Kou98,Whe00,Tho04}. Magnetars have provided popular models to explain some of the most mysterious phenomena in the universe, such as fast radio bursts and gamma-ray bursts,  although their origin remains unclear. For example,  a magnetar with a magnetic field $\geq 10^{16}$ G and a spin period $\leq 1$ ms, its releases the spin-down energy at $\sim 10^{51}$ erg sec$^{-1}$ in the first 20 sec after its formation,  and likely drives a collimate jet and produces gamma-ray bursts \citep{Met11,Met15}.  If a magnetar with a magnetic field of $10^{14} - 10^{15}$ G and a spin period of $\sim 1-10$ ms radiates away its spin-down energy following a dipole formula,  its luminosity becomes much less but lasts for much longer.   \refcite{Mae07},  \refcite{Woo10}, and
\refcite{Kas10} showed that SNe containing magnetars could power exceptionally luminous transients to explain SLSNe \citet{Qui11,Gal12,Ins13} because a fraction of the magnetar's spin energy is emitted as optical radiation at a later time when the ejecta is becoming optically thin.  Recent SLSNe observations by  \refcite{Ins16,Ins17,Mar18,Nic19,Mar19} agreed with light curves models of magnetar-powered SNe by \refcite{Kas16,Mor17,Des18,Des19}. 

\begin{figure}[pb]
	\centerline{\psfig{file=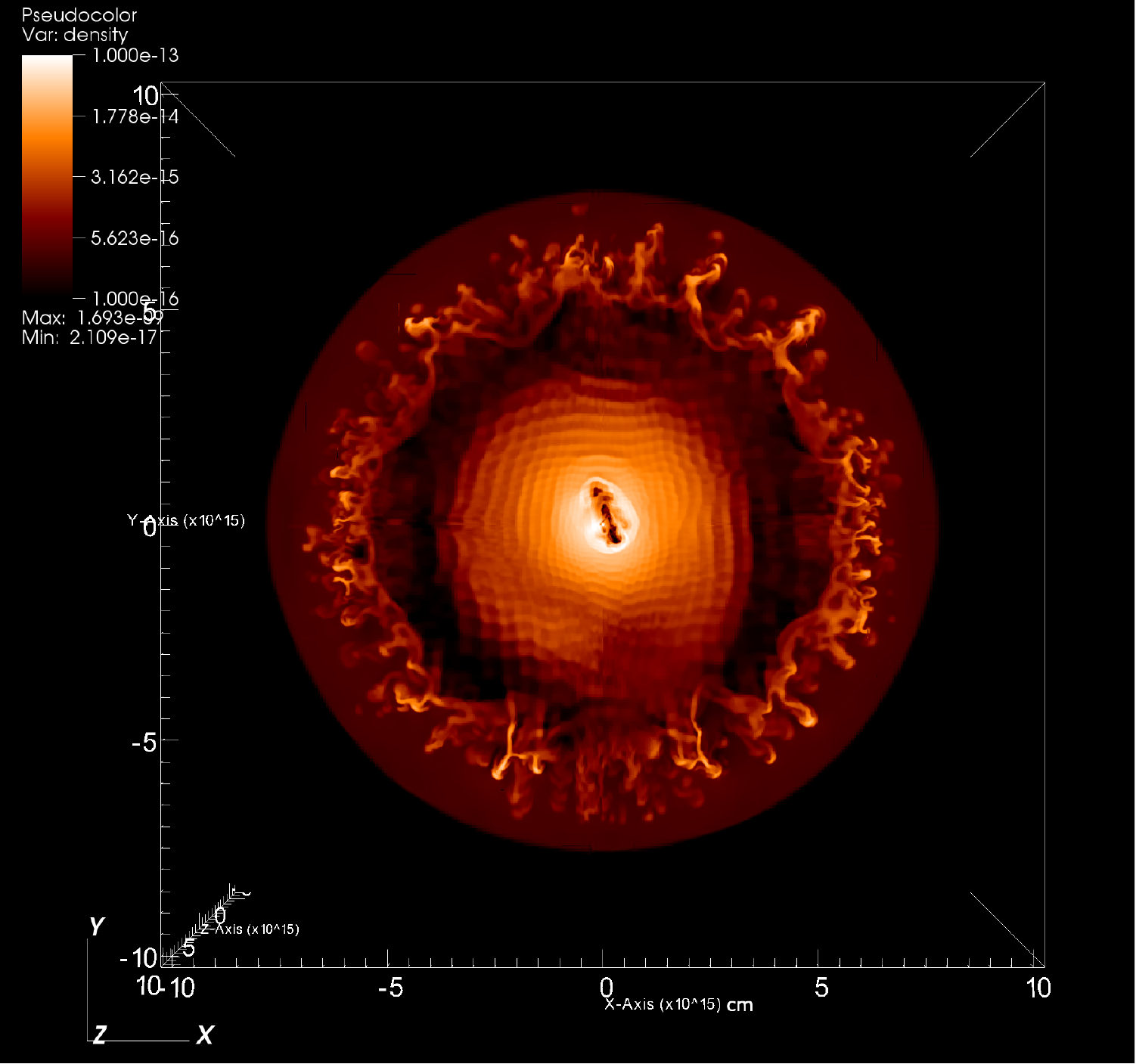,width=12cm}}
	\vspace*{8pt}
	\caption{ A magnetar-powered SLSN in the nebula phase. At this moment, SN  ejecta has expanded to $10^{16}$ cm similar to the size of our solar system. Large-scale fluid instabilities occur in ejecta and lead to a strong mixing, which affects the spectra and light curves of SLSNe. \label{f1}}
\end{figure}

These models successfully produced light curves in a good fit with SLSNe; however, a critical drawback exists among them due to the limitation of one-dimensional calculations.  The matter accelerated by the magnetar's radiation affects the SN dynamics when the magnetar's total deposited energy is comparable to the explosion energy.  The accelerated matter quickly piles up to form a dense and thin shell seen in previous 1D models. This shell structure looks like a spike in the density profile, indicating much mass accumulated in a shell of volume $4r_s^2dr$ at the shell radius, $r_s$ with $dr << r$, the radius of SNe ejecta. If the formation of this shell in 1D is real,  radiation should be trapped inside this shell, and the resulting spectra would show most of the ejecta moving at a single speed, which is not supported  by observation. What is the origin of this shell in 1D? The thin shell indicates the accumulating matter in a small volume due to fluid instabilities, which require one fluid to pass over the other and cannot be modeled in 1D. Therefore, a second or third dimension is needed to simulate fluid instabilities and evolve the consequent mixing. Furthermore, similar to the magnetar-powered SLSNe, early studies of pulsar wind nebulae by \refcite{Che82,Che92,Jun98,Blo01,Che11} suggested the collisions of pulsar winds to their surrounding CSM are subject to hydrodynamic instabilities. Magnetar's wind within a young SN remnants were investigated by \refcite{Che16,Kas16,Blo17,Suz17,Suz18} with two-dimensional simulations. Recently, \refcite{chen20a} carried out a 3D simulation to study the turbulent mixing in a magnetar-powered SLSN . These 
multidimensional simulations suggested that the fluid instabilities are  strong enough to break the accelerated shell into filamentary structures and cause a large-scale mixing in the SN ejecta, as shown in Figure \ref{f1} modified from \refcite{chen20a}. 

\subsection{Pulsational Pair-Instability SNe}

\begin{figure}[pb]
	\centerline{\psfig{file=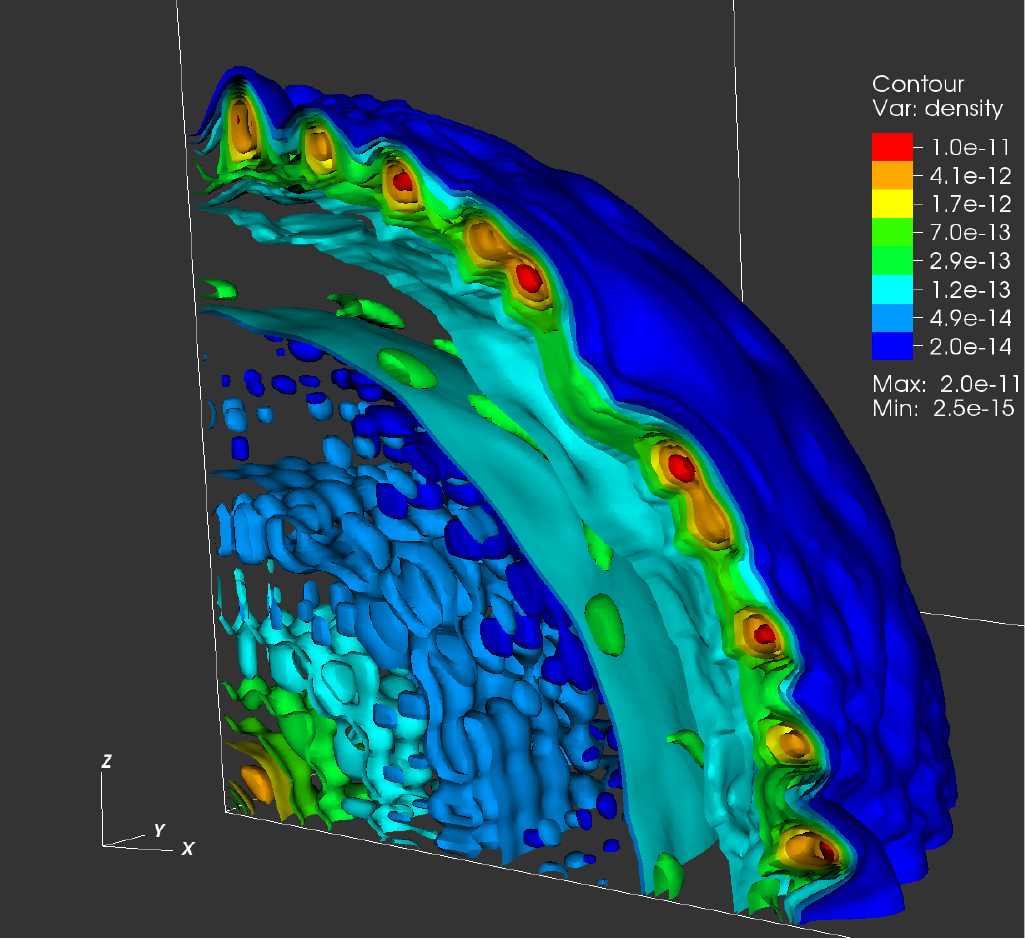,width=12cm}}
	\vspace*{8pt}
	\caption{In their final years before exploding, massive stars undergo repeated eruptions due to the PI. Dense clumps (red nuggets) appear at the outer regions of emerging shells,  $\sim 10^{16}$ cm from the progenitor star located at the left-bottom corner.  \label{f2}}
\end{figure}

SNe of massive stars can explain SLSNe through two different channels.  Before a massive star of 80 - 150 \Ms\ dies,  its core temperature reaches $> 10^9$ K, making energetic photons $\ge 1 Mev$ in the tail of Maxwellian distribution convert into electron-positron pairs through the photon-nuclear collision. The pair-production removes radiation pressure from the core and triggers a dramatic contraction. This contraction increases the core's temperature and density, which then leads to violent carbon burning and triggers several eruptions pulsationaly \citet{barkat1967,Woo18}.  
Energy and mass of eruptions increase as the mass of a star rises. 
The first strong eruption likely removes all of the  hydrogen envelope at once and yields either a faint Type IIp SN. Subsequent eruptions later overtake and collide with the ejected hydrogen envelope and produce a luminous Type IIn SN.  Especially, massive stars of $100 - 130$ \Ms\ are promising candidates to become SLSNe because their characteristic time scale between PI eruptions is several years, and the collisions between ejected shells of speeds $\sim$ 1000 km s$^{-1}$ happen at $r \sim $ 10$^{15} - 10^{16}$ cm \citep{Woo18}. The collision energy can reach several 10$^{50}$ erg, which is mostly dissipated into optical light and results in a luminous transit, known as "pulsational pair-instability supernova'' (PPI SN). \refcite{woosley2007} modeled the PPI SN of a 110 \Ms\ star with the 1D stellar evolution code, \KEPLER{} \citep{kepler,heger2001}, and calculated the light curve with the \STELLA{} code \citep{stella} to explain the first SLSN, SN 2006gy \refcite{smith2007a}.  Their model produced $10^{50}$ erg of radiation, ten times more than a typical CCSN and demonstrated PPI SNe as candidates of SLSNe. However, the catastrophic shell collisions of ejecta in PPI SNe are subjected to the Rayleigh-Taylor Instabilities (RTI) and lead to a dense shell formation, that is a similar issue in the previous 1D magnetar-powered SNe. 
This density shell caused large fluctuations in the bolometric light curves; similar findings were also confirmed in \refcite{wet13d}. Recently, \refcite{chen14a} presented the two-dimensional hydrodynamics simulations of PPI SNe and showed this 1D  dense shell transforming into RTI fingers in 2D. At the end, these RTI fingers evolved to a large-scale mixing in the regions where
 the PPI SN radiation originates. The latest 2D and 3D radiation-hydrodynamics simulations by \refcite{chen19} revealed the evolution of fluid instabilities under the effect of radiative cooling and its impact on the SN light curves.
Figure \ref{f2} shows the 3D structure of a PPI SN from \citep{chen19};  radiative cooling transforms the RT fingers into dense clumps, which are also hot spots of radiation.

\subsection{Pair-Instability SNe}

For a  massive star of 140 - 250 \Ms, the pair-instability happens during the central oxygen burning. Unlike the PPI SNe, the core contraction becomes runaway, and ignites the explosive oxygen burning that blows up the entire star as  a pair-instability supernova (PI SN). The ideal of PI SN were first introduced by \citet{bark1967}, further developed by \citet{ober1983,glatzel1985,stringfellow1988,heger2002,heger2010}. PI SNe produce $10^{52} - 10^{53}$ erg of explosion energy and synthesize $0.1 - 50$ \Ms\ of \Ni,  10 --100 times larger than normal SNe.  The physics of PI SNe is more robust than other types of SNe due to the better understanding of thermonuclear  physics \citep{heger2002,scannapieco2005,candace2010-2}. Furthermore,  a large amount of \Ni\ decay can release  enormous radiation, which 
naturally makes PI SNe excellent candidates for SLSNe. The observational signatures of PI SNe  predicted by \refcite{kasen2011,Koz14,Koz15,Wha14} suggested that PI SNe will be visible in the near-infrared (NIR) at $z \sim 10-15$ for the next-generation space telescopes to probe the masses of first stars \citet{Har18,Tak18,chen17b}.

\begin{figure}[pb]
	\centerline{\psfig{file=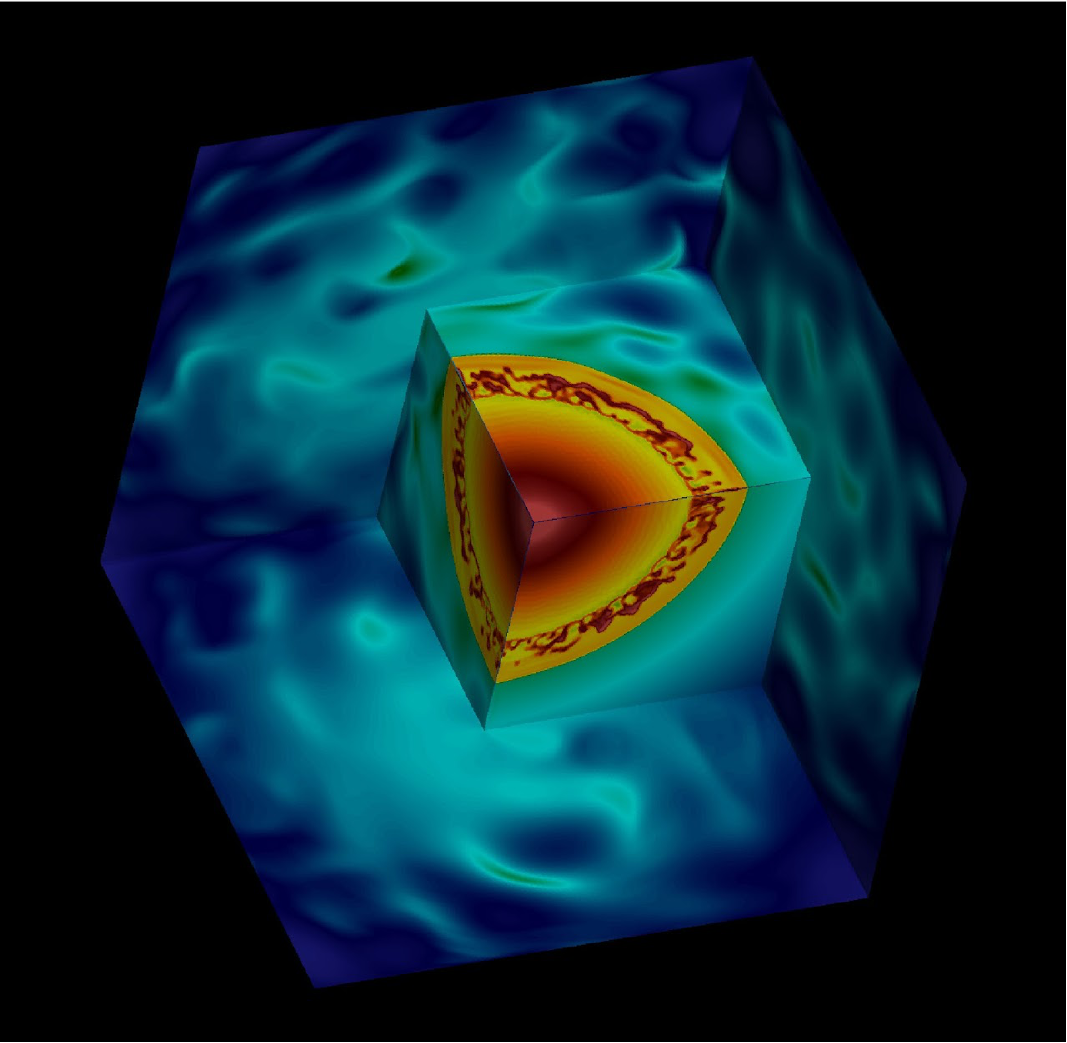,width=12cm}}
	\vspace*{8pt}
	\caption{Pair-instability supernova in a 3D box. Hot-cool colors represent the density from high to low. Unlike the magnetar-powered and PPI SNe, only minor instabilities are found in the core (red-yellow variations).  \label{f3}}
\end{figure}

Recent PI SNe models by \refcite{Cha15} suggested that the stellar rotation can shift to a lower mass limit of the progenitor stars from 140 to 85 \Ms\,, and  \refcite{chen2015} found that  \Ni\ production decreases significantly in PISNe of rapidly rotating stars.  Sophisticated multidimensional models of PI SNe by \refcite{chen2011,chen14a,Gil17} suggested that the fluid instabilities in PI SNe  of blue supergaints are too weak to break the spherical-symmetry of ejecta; only those of red supergiants show visible mixing as shown in Figure \ref{f3} taken from \refcite{chen14d}.  Therefore, the results of earlier 1D calculations remain valid. On the other hand, tens of \Ms\ Ni\  decay can release several $10^{51}$ erg of energy, which heats up the inner ejecta. 
Like the magnetar,  the matter accelerated by the decay energy piles up a shell and traps the radiation within it, known as the \Ni\ bubble issue found in the 1D radiation hydrodynamical calculations by \refcite{Wha14,Koz15}. Recently, \refcite{chen20b} performed long-term 2D simulations of PI SNe to investigate the \Ni\ bubble issue and found that the energy from \Ni\ decay can make a broad shell in the inner ejecta but drives no mixing. However, $\sim 30\%$ of decay energy is used to accelerate this shell instead of luminosity; thus, the PI SNe become fainter.

\section{Toward Realistic Models of Superluminous Supernovae and Future Perspectives}

Understanding the physics of SLSNe holds the key to properly identify them as 
more of them are now discovered by new SN surveys, such as the {\it Palomar Transient Factory}
\citep{ptf1} and the {\it Panoramic Survey Telescope and Rapid Response System} \citep{panstarrs}.   
So far, magnetar-powered and pair-instability SNe (both kinds) are promising candidates for SLSNe 
because their emission mechanisms naturally explain the unusual luminosity of SLSNe. The physical properties of SLSNe models are briefly summarized in Table \ref{tab1}.
Light curves of these models have good accordance with SLSNe observations. However, none of these models can fully match up with the observed spectra. With the advance 
of modern supercomputers, realistic 2D and 3D SLSNe simulations reveal the importance of fluid instabilities, which was a hurdle in the 1D models. Fluid instabilities in magnetar-powered and PPI SNe  models break down the spherical symmetry of the ejecta and rearrange their elemental abundances, and that profoundly affects the resulting spectra. Based on the 2D and 3D simulations of PI SNe  by \refcite{chen2011,chen2014a,chen20b}, the major mixing in PI SNe is driven by  the reverse-shock  formed in the red supergiant progenitor stars. Heating from the \Ni\ decay only causes a broad shell at the boundary of the inner ejecta without contributing much mixing. Therefore, the mixing in PI SNe is weak, and their spectra predicted from previous 1D calculations remain valid. In the case of PPI SNe, mixing \Cx\ and \Ox\ without the \Fe\ group elements in its ejecta leaves a distinctive feature on the spectra \citet{chen2014,chen19}. Finally,  the magnetar engine is the most popular model for observers, and it is widely used to fit many light curves of SLSNe. Unlike PI SNe and PPI SNe, the mixing in the magnetar-powered SNe is much stronger due to a large energy injection from the magnetar \citet{chen16} which accelerates the iron-group elements to a speed of $\sim$ 10,000 km$\,$s$^{-1}$, resulting in the broad-line features in the spectra predicted by \refcite{chen20a}.  Although multidimensional hydrodynamic simulations shed light on the mixing in SLSNe, more sophisticated models are required before predicting realistic signatures of SLSNe. To improve existing models,  we have to develop 3D radiation hydrodynamics simulations with relevant atomic physics to properly catch the turbulent mixing of ejecta and to calculate the consequent luminosity. In addition, the scheme of energy injection from a magnetar can be improved based on the GR-MHD simulations of magnetar formation \citet{Mos15}. Furthermore, post-processing the simulation results with the time-dependent monte-carlo radiative
transfer code such as SEDONA \citet{Kas10} to obtain the detailed spectra will provide powerful observational diagnostics of models and their explosion mechanisms.

\begin{table}[ht]
	\tbl{Models of Superluminous SNe}
	{\begin{tabular}{@{}lll@{}}	 \toprule
			Mass of progenitor star [\Msun]	&	  Explosion mechanism  & Luminosity source  \\ \colrule
			$25\hphantom{0} \quad\leftrightarrow\quad 50$ &  Core Collapse SNe & Magnetar spin-down \\
			$80\hphantom{0} \quad\leftrightarrow\quad 150$ & Pulsational Pair-Instability SNe & Collisions of ejected shells\\
			$150 \quad \leftrightarrow\quad 250$ &  	Pair-Instability SNe & \Ni\ decay \\  \botrule
	\end{tabular}}
\label{tab1}
\end{table}

The progenitors of SLSNe and their origins remain uncertain; we believe they are related to deaths of very massive stars; $25 - 50 \Ms$ for magnetar-powered SNe,  $80 - 150 \Ms$ for PPI SNe, and  $150 - 250 \Ms$ for PI SNe.  
To optimize the detection rates of SLSNe, observers should look for the sites of massive star formation.
 \refcite{humphrey1979,davidson1997,r136} suggested the formation of massive stars $> 100 \Ms$ in our galactic center, and modern cosmological simulations \citep{hir13,hir14}  suggested that the mass scale of the first stars would range from tens to hundreds of \Ms. These results suggest that SLSNe may occur in both the local and early universe.  The discovery of SLSNe in the local universe implies that 
such events might be visible at high redshifts  and demonstrate SLSNe as a promising probe to the early universe. For example,  SN 2007bi at $z =$ 0.127 \citep{2007bi} and SN 2213-1745 at $z =$ 2.06 \citep{cooke12}, and SN 1000$+$0216 at $z =$ 3.90 \citep{cooke12} and SN 2006oz at $z =$ 0.376 \citep{lel12,wet13d}  shall be visible to the {\it James Webb Space Telescope} ({\it JWST}) and the {\it Nancy Grace Roman Space Telescope} at $z \sim$ 20 \citet{kasen2011,pan12a,wet12b,wet12a,wet12e,hum12}. 
The next-generation of ground telescopes, such as the {\it Thirty Meter Telescope} (TMT), the {\it European Extremely Large Telescope} (E-ELT), and the {\it Giant Magellan Telescope} (GMT), will provide the most detailed spectra for SLSNe. Therefore, gaining a complete understanding of SLSNe is timely and essential. With the advance of models and observations, we soon will consolidate the physics of SLSNe and use their spectacular light as a standard candle to probe the properties of the first stars\citep{abel2002,turk2009,hir14}, the first supernovae \citep{candace2010,wet12c,wet12d,jet13a}, and the first galaxies \citep{fsg09,glov12,dw12,fg11}. Doing so will further unveiling the early 
cosmological reionization and chemical enrichment \citep{whalen2004,mbh03,ss07,bsmith09,ritt12,ss13}, as well as the origins of supermassive black holes \citep{vol12,woods19}.

\section*{Acknowledgments}
The author thanks Stan Woosley for many useful discussions.  This work is  supported by the Ministry of Science and Technology, Taiwan, R.O.C. under Grant no. MOST 107-2112-M-001-044-MY3. KC thanks the hospitality of the Aspen Center for Physics, which is supported by NSF PHY-1066293, and the Kavli Institute for Theoretical Physics, which is supported by NSF PHY-1748958. Our computational resources are provided by the National Energy Research Scientific Computing Center (NERSC), a U.S. Department of Energy Office of Science User Facility operated under Contract No. DE-AC02-05CH11231,  the Center for Computational Astrophysics (CfCA) at National Astronomical Observatory of Japan (NAOJ),  and the TIARA Cluster at the Academia Sinica Institute of Astronomy and Astrophysics (ASIAA).

\bibliographystyle{ws-ijmpd}

\end{document}